# Cooperation in the Gig Economy: Insights from Upwork Freelancers

ZACHARY FULKER, Northeastern University, USA
CHRISTOPH RIEDL, Northeastern University, USA

Existing literature on online labor markets predominantly focuses on how freelancers *individually* complete tasks and projects. Our study examines freelancers' willingness to work *collaboratively*. We report results from a survey of 122 freelancers on a leading online labor market platform (Upwork) that examine freelancers' preferences for collaborative work arrangements, and that explore several antecedents of cooperative behaviors. We then test if actual cooperative behavior matches with freelancers' stated preferences through an incentivized social dilemma experiment. We find that respondents cooperate at a higher rate (85%) than reported in previous comparable studies (between 50-75%). This high rate of cooperation may be explained by an ingroup bias. Using a sequential mediation model, we demonstrate the importance of a sense of shared expectations and accountability for cooperation. We contribute to a better understanding of the potential for collaborative work on online labor market platforms by assessing if and what social factors and collective culture exist among freelancers. We discuss the implications of our results for platform designers by highlighting the importance of platform features that promote shared expectations and improve accountability. Overall, contrary to existing literature and predictions, our results suggest that freelancers in our sample display traits that are more consistent with belonging to a coherent group with a shared collective culture, rather than being anonymous actors in a transaction-based market.

CCS Concepts: • **Human-centered computing** → **Computer supported cooperative work**; *Empirical studies in collaborative and social computing*.

Additional Key Words and Phrases: Macrotask crowdsourcing, crowd teams, cooperation experiment

## 1 INTRODUCTION

Online labor markets offer on-demand access to freelancers from around the world who can complete a variety of tasks. Many of them are specialized on knowledge intensive professional work: tasks that require sophisticated skills like programming, graphic design, and data science. Consider Upwork, a leading online labor market platform, which has a gross services volume of over $3.3 billion, over 16 million freelancers from around the world, and is used by 30% of Fortune 100 companies [17]. Despite their growing size, a limitation shared by Upwork and similar online labor markets facilitating knowledge-based crowd work, has been their focus on individual work. Many problems, however, require interdependent contributions from multiple workers that cannot be fully, or easily, decomposed into independent individual tasks [61]. Strengthening collaborative work on online labor markets could also offer a viable pathway to make such work more sustainable

Authors' addresses: Zachary Fulker, fulker.z@northeastern.edu, Northeastern University, 360 Huntington Ave, Boston, MA, USA, 02115; Christoph Riedl, Northeastern University, 360 Huntington Ave, Boston, USA, c.riedl@northeastern.edu.

and engaging by addressing negative effects of social isolation experienced by many "gig workers" [26, 68]. Therefore, there is a need for platform theory and design that can facilitate team-based crowd work [38, 39, 57]. Our study focuses on knowledge-based crowd work which we define as work that is knowledge intensive (rather than capital intensive), requires specialized skills, is relatively long-term (hours, rather than minutes), and leans toward a more complex spectrum (compared to pure "microtask" crowdsourcing that is unlikely to benefit much from collaborative work). Examples of such knowledge-based crowd work include programming, design, translation, and creative writing. Throughout, we use the term "freelancer" to refer to the category of crowd and "gig" workers who engage in those knowledge-based work through online labor markets. We contribute to this research stream by focusing on an important antecedent of collaboration: freelancers' cooperativeness and interest in collaborative work.

Several proof-of-concepts developed in the CSCW community demonstrate both the feasibility and benefits of team-based crowd work (e.g., [54]). However, team-based crowd work can only be successfully deployed at scale if freelancers are interested in collaboration and willing to cooperate with other freelancers. Therefore, evidence is needed on whether there is widespread interest in collaborative work among freelancers to shed light on whether collaborative crowd work is likely to succeed. The answer to this question is not obvious since freelancers come from different regional backgrounds (and hence have no shared cultural context), typically interact only through virtual channels, and often compete with each other for the most lucrative jobs. In this paper, we investigate individual traits and behaviors related to cooperation and collaboration in a diverse sample of Upwork freelancers. We address four key research questions: (1) are freelancers interested in team-based, collaborative work? (2) does interest in collaborative work vary across geographic regions, full vs. part-time work, posted hourly wage, and technical domains? (3) does freelancers' actual cooperative behavior match with their stated preferences? and (4) what are the antecedents of cooperative behavior that could inform platform design? We conduct a survey of 122 freelancers on the Upwork platform to assess freelancers' desire to collaborate with fellow freelancers, test known antecedents of cooperation identified by organizational behavior research, and examine freelancers' willingness to cooperate in an incentivized social dilemma. The social dilemma experiment allows us to capture the choice between acting in a selfish way that benefits oneself but harms others, and maximizing shared benefits [18].

We find that most of our respondents are interested in collaborating with fellow freelancers (82% hold a net positive view on collaborative work with other freelancers), and that this desire persists across key demographic categories. We also find support for a range of factors previously shown to relate to cooperation in traditional organizations. Most strikingly, we find extremely high rates of cooperation (85%) in our incentivized social dilemma. We theorize that ingroup effects are a likely explanation. Finally, we explore antecedents to successful collaboration through a sequential mediation model using structural equation modeling (SEM). This demonstrates the importance of a sense of shared expectations and accountability in fostering cooperation among freelancers. This result provides platform designers with a potential pathway to strengthen cooperative norms. Overall, the respondents to our survey display traits that are more consistent with belonging to a coherent group with shared collective culture, rather than being anonymous actors in a transaction-based market.

We expand prior work studying freelancers' perceptions of their work environment (e.g., [11, 23, 29, 64, 68]) with insights into their desire to engage in more collaborative work, and their level of actual cooperative behavior. While prior work has shown that freelancers collaborate with each other outside the actual task work (e.g., to find lucrative tasks and manage administrative overhead; [29]), and collaborate with each other in specific proof-of-concept instances (e.g., [57, 67]), we show that freelancers in our sample have a general desire for collaborative work. Our findings offer a mandate to platform designers to strengthen their support for team-based crowd work by calling out the desire of freelancers for collaborative work arrangements. If successful, team-based crowd work could improve the quality of work available on online labor platforms by providing these freelancers with an option for more social, cooperative, and meaningful work.

Our findings that behavior in online labor markets can diverge in important ways from the logic of similar online markets for goods, expands upon previous crowd work research contrasting client-freelancer relationships from buyer-seller relationships [51]. For example, freelance workers behave more like parties to an employment contract rather than sellers in a competitive market [16]. In crowd work, scholars have suggested that a lack of shared culture will have broad implications for how work needs to be managed and organized [41]. Our work suggests there is a surprising amount of shared culture and possibility (and desire) for teamwork. This contributes to a recent call for a deeper assessment of the relational dynamics within online labor markets [51]. Further, our work strengthens the claim that viewing online labor platforms as anonymous, transaction-based markets is not consistent with freelancer intentions and behaviors.

In summary, we make three main contributions.

- Evidence of the willingness of freelancers on a major labor platform to cooperate, as well as their stated desire to participate in team-based collaborative work. Such desire to collaborate is consistent across freelancers from different countries, technical & non-technical work, and full vs. part-time status.
- The observed high rate of cooperation of 85% appears to be best explained by ingroup bias in which freelancers identify with other members of the same online labor market.
- A sense of shared expectations and accountability influences trust in fellow freelancers which is an important antecedent of cooperation among freelancers. This provides a pathway for platform designers to strengthen cooperative norms, and has implications for organizations that rely on external contributors.

## 2 RELATED WORK

Solving more complex problems through crowd work requires cooperation between freelancers [30]. In short, they need to form a team in order to facilitate shared effort, coordinated work, and adaptation to unforeseen challenges [53]. In addition to expanding crowd work to address more complex problems, collaborative work arrangements also promise higher output quality [8], and social benefits from interpersonal interactions and social support [28, 68]. We build on related work in three areas to identify our research gap and inform our research design: work on team-based crowdsourcing, offline temporary teams, and behavioral theory that has explored antecedents of cooperation in traditional organizations.

### 2.1 Crowdsourcing and Crowd Teams

Surveys and other ethnographic studies have started to shed light on attitudes, behaviors, and perceptions of (individual) crowd and so-called "gig" workers. For example, [11] study motivation to participate in (individual) crowd work, [23] show how task characteristics shape perceptions of crowd work, [64] study how much time freelancers spend on administrative tasks, and [29] show freelancers use off-platform social networks to fulfill technical and social needs unaddressed by the platform they work on. Autonomy has been shown as one of the most salient values held by freelancers and a primary motivator for their participation [19]. This could suggest that widespread adoption of collaborative work may be unlikely despite the successful deployment of various proof-of-concept implementations of collaborative crowd work systems.

Several prior empirical studies have examined specific instances of team-based crowdsourcing. For example, Riedl & Woolley [57] look at the relative importance of incentives, skill, and emergent processes for team performance, and Blasco et al. [10] study the ability of crowd teams to self-organize. Several other studies exist (e.g., [20, 34, 50]). While these studies demonstrate that freelancers can successfully collaborate and often develop sophisticated emergent collaboration processes to be successful, participants in these studies were often motivated by powerful incentives and/or shared interests, and were aware that the projects were research experiments. Whether the broader population of freelancers is interested in collaborative crowd work remains unclear.

Besides empirical studies that employ teams, a large body of work in the CSCW community has focused on designing systems that support collaborative crowd work (e.g., [39, 61, 67]). Rather than rely on team work with (synchronous) collaboration, this work has often tried to avoid (or limit) interdependencies among workers by focusing on decomposing complex tasks into smaller subtasks, which are then assigned to individual workers, and finally results are reassembled upon completion [38]. These systems often organize workers using workflows [1, 7, 43, 54]. Most often, workflows are static with a predefined set of tasks and a sequence in which these tasks are computationally assigned, completed, and reassembled [43]. While such static workflows have proven to be effective at solving large-scale problems with predictable interdependencies and contingencies [7, 42], knowledge work often follows unpredictable paths. In response to the challenge of crowdsourcing complex knowledge work, adaptive workflows and clearly defined roles have been proposed as possible solutions [39, 53, 61, 67]. This, however, would require more explicit collaboration among freelancers. Successful adaption to unforeseen challenges is dependent on a team's ability to coordinate and collaborate in identifying, proposing, and approving workflow changes. Likewise, individualized roles with little prespecification of work encourages adaptation but greatly increases the need for emergent coordination [53]. This places the burden of successful coordination and collaboration back on freelancers. It is unclear if freelancers are interested in such collaborative work arrangements, whether they are willing to shoulder the necessary cooperation, and whether workflows or roles are more desirable to accomplish it.

## 2.2 Organizational Theories of Temporary Teams, Cooperation, and Job Satisfaction

*2.2.1 Offline Temporary Teams.* Related work on offline temporary teams can provide insight into the emergence of effective cooperation and coordination outside of traditional work settings with established interpersonal relationships. Temporary teams, unlike ongoing teams, have a finite goal and the expectation the team will be dissolved after task completion [60]. Ethnographic work has detailed the behaviors that enable temporary teams to coordinate, for example, on film crews [5] and in healthcare [45, 66]. That research has emphasized the importance of *shared expectations* and *accountability* in facilitating coordination and maintaining cooperation. Unlike ongoing teams where past interactions establish norms, and future interactions increase accountability between members, temporary teams lack these in-built mechanisms to support norms and accountability.

In addition to these challenges, temporary crowd teams may face additional complications. First, the client organization for whom work is to be conducted changes frequently, preventing free-lancers from developing and learning about the norms associated with a single client organization. Second, since freelancers are individuals drawn from a global pool, there may be even less shared understanding and culture among freelancers to help facilitate coordination and cooperation. Third, freelancing is often seen as solitary and autonomous work. Individuals who self-select to work as freelancers could simply be less interested in collaborative work arrangements because they have chosen this form of work precisely for that reason. In summary, it is unclear if temporary crowd teams can consistently congeal into productive units, and thus whether concepts like shared expectations and accountability shown to matter in offline temporary teams also extend to temporary crowd teams.

*2.2.2 Theories of Cooperation.* The extensive organizational behavior literature on collaboration also informs our work. Interdependent tasks that require coordination across multiple individuals are a core part of professional work [30, 44]. Willingness to cooperate is an obvious but easily overlooked precondition to effective collaboration in teams. Existing theory has established important antecedents for cooperation in organizational settings. This includes perspectives focused on relational [63], task [22], organizational [62], and personal factors [15]. Empirical research has sought to quantify the relative impact of each of these factors on realized cooperation. Results have shown that each of these antecedents exerts a significant and independent influence on cooperation [71]. This suggests that no single factor fully explains cooperation in isolation. Rather, cooperation is best understood through a holistic view of the environment in which it takes place.

Freelancers are geographically dispersed strangers who are often competing with each other for the most attractive tasks in an online market. It would therefore seem to be a hostile environment for cooperation. When workers are distributed, cultural differences and coordination are particularly challenging and may prevent cooperation [33]. Currently, freelancers on online labor markets have limited interactions with other freelancers and are hence often thought of as being independent [29] and transaction-focused rather than belonging to a shared organization. Therefore whether the factors documented to support cooperative behavior in organizations translate to crowd work settings is unclear.

*2.2.3 Job Satisfaction and Relational Job Design.* While the degree of viable cooperative behaviors between freelancers is not clear, existing literature on job satisfaction and relational job design suggests these workers may benefit from the introduction of collaborative work. Even before the advent of online labor market platforms, freelancing presented a trade-off between community and autonomy. In a 2008 article, one programmer lamented "I could either have a job, which gave me structure and community, or I could be freelance and have freedom and independence. Why couldn't I have both?" [25]. The relational architecture of jobs (i.e., with whom and how you communicate while carrying out your work) can provide social support, as well as connections to those impacted by your work, increasing worker motivation and well-being [26]. Such support and connections are often lacking in "gig" work [68]. Another factor of job satisfaction where crowd work can fail to match traditional employment is task significance, or how important and necessary a task seems to the worker completing it [31]. Because of their status as external contributors to tasks requested by client organizations, freelancers may have limited ability to see and understand how their work contributes to the larger problems being addressed. Experiments have shown that exposure to cues about task significance can enhance job performance by promoting a deeper understanding of the value of one's work [27]. Participating in collaborative crowd work could increase job satisfaction and performance among freelancers by improving the relational architecture of their daily work, thereby increasing motivation and positive work behaviors. Due to these potential improvements to job satisfaction, many freelancers may be interested in engaging in collaborative work.

# 3 METHODS

We created a short, IRB approved survey (∼ 7 minutes est.) to measure factors related to cooperative behavior. Most were measured using multiple Likert-style items adapted from previous publications with a few exceptions. To assess relational factors, we measured trust in fellow freelancers [71] and identification with the broader freelancer community [71]. Because the freelancers we surveyed currently work on a variety of different individual tasks without clear interdependencies, we do not consider the role of any task specific properties in cooperation. As a measure of organizational factors, we collected responses on intrinsic and extrinsic job satisfaction [71]. To assess the personal characteristics of respondents, we measured personal cooperativeness [71] and goal orientations [70]. Additionally, we developed a set of questions to assess if respondents believed freelancers had a shared sense of acceptable behaviors with regard to the completion of their work, and a shared sense of acceptable behavior by requesters. Furthermore, we asked if respondents believed freelancers who engaged in unethical behaviors would be held accountable. Finally, we measured respondents' desire to participate in team-based collaborative work, and their preference for how such collaboration should be facilitated.

Importantly, we also measured actual cooperative behavior. Respondents participated in an incentivized prisoner's dilemma at the start of our survey in which they were told they would be paired with a fellow freelancer, and could earn a bonus of between $0 and $2 (representing an additional 33% of respondent compensation) based on their response. In any collaborative work arrangement, individuals often must choose between acting in a selfish way that benefits themselves but harms collaborators, or maximizing shared benefits. This scenario is modeled by the class of games known as social dilemmas in game theory. For example, consider the common scenario where a worker could choose to reduce their effort and free-ride off the work of others, or they could choose to contribute to the work equally. The stability and quality of any cooperative work arrangement is dependent upon workers not making the selfish choice. Therefore we presented respondents with an incentivized social dilemma to test their behavior. Such a social dilemma has been used as a proxy of the collective efforts required to achieve collective goals and interests in previous research [18]. The full set of survey questions is listed in the Appendix.

We recruited survey participants from Upwork in May of 2022. We invited participants to take part in our survey in the order their profiles appeared as suggestions from the Upwork algorithm. Only invited participants could take part in the survey. To create a sample of freelancers on Upwork with diverse skills and backgrounds, we recruited freelancers from five common job categories: backend development, data science, accounting, graphic design, and business writing. Respondents were paid $6 for completing our survey plus any additional bonus resulting from their answer to the social dilemma (average payment comes out to about $60 per hour). The response rate among invited participants was about 25%. We took the following steps to further increase the representativeness of our sample. Given that some types of freelancers—for example those with lower posted hourly wages—may be more likely to respond than others, we also collected the profile data of all invited participants including, but not limited to, measures of experience, earnings, and location. We find significant differences in the posted hourly rate and location of those who chose to respond to our survey (Fig. 1). This suggests that raw survey responses are not representative of all freelancers on Upwork—i.e., we find some sample bias. To correct for this bias, we use the inverse of the predicted probability to respond to our survey from a logistic regression to reweight our survey data. We report the unadjusted descriptive statistics below, and report sample weight adjusted modeling results in Section 5. Overall, reweighting our survey does not meaningfully change any of our key results, but this method is a simple and effective way to reduce bias in survey research where population-level demographic data can be collected.

We collected 122 survey responses from Upwork freelancers. Respondents came from 66 different countries and had a posted hourly wage of $35 on average (ranging from $5 to $350). The most common age group was 25-34 years old (52%) followed by 35-44 years old (25%). Respondents were 72% male. Almost half (49%) of respondents were experienced platform freelancers reporting more than three years of experience, and only 13% had used the platform for less than a year. The majority of respondents (59%) reported not using any other online labor market besides Upwork. Likewise, a majority (65%) of respondents earned over half of their income from freelance work.

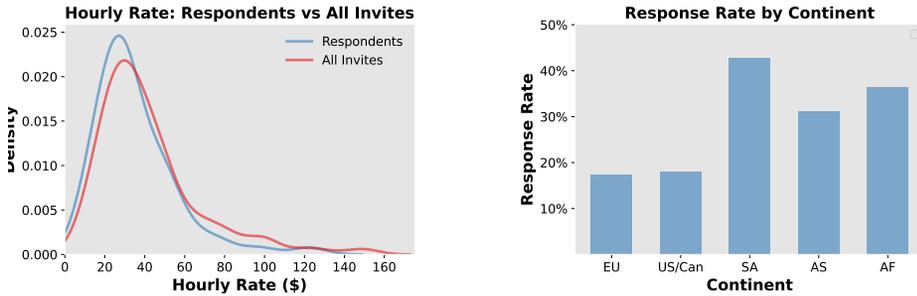

Fig. 1. We used the profile data of invited respondents to test for response bias in our data. We found a significant difference in the mean posted hourly rate of respondents who completed the survey when compared to the set of all invited respondents. We also found a significant difference in the proportion of invited respondents who completed the survey (i.e. response rate) by continent.

Table 1. Averaged responses across all survey respondents on a 1-Strongly Disagree to 5-Strongly Agree Likert scale, unless otherwise noted. Cronbach's alpha gives a measure of internal consistency for measures consisting of multiple survey items.

| Scale | Mean | Std.Dev. | Cronbach's Alpha |
|---|---|---|---|
| Intrinsic Satisfaction (5 items) | 3.42 | 0.67 | 0.80 |
| Extrinsic Satisfaction (4 items) | 3.11 | 0.79 | 0.80 |
| Trust in Fellow Freelancers (4 items) | 2.58 | 0.70 | 0.71 |
| Shared Expectations & Accountability (3 items) | 3.08 | 0.72 | 0.73 |
| Identification w/ Freelance Community (4 items) | 3.34 | 0.86 | 0.90 |
| Personal Cooperativeness (5 items) | 2.65 | 0.64 | 0.68 |
| Learning Goal Orientation (5 items) | 3.51 | 0.65 | 0.86 |
| Prove (Performance) Orientation (4 items) | 3.14 | 0.72 | 0.69 |
| Avoid (Performance) Orientation (4 items) | 2.44 | 0.91 | 0.79 |
| Desire to Collaborate (4 items) | 2.99 | 0.96 | 0.89 |
| Predict others will Cooperate | 64.18% | 22.97 | - |
| Social Dilemma | 85% (Cooperate) | - | - |
| Prefer Workflow or Roles | 54% (Roles) | - | - |

The freelancers surveyed were also highly educated with 41% having a bachelor's degree and 51% having a graduate degree.

## 4 DESCRIPTIVE RESULTS

The descriptive results of our survey demonstrate three main findings: (1) Freelancers are interested in collaborative work arrangements; (2) Factors that support cooperation in traditional organizations are also present among Upwork freelancers; and (3) Freelancers will cooperate in a real social dilemma. In Section 5, we use a structural equation model (SEM) to examine what relationships exist between the factors most predictive of cooperation in our social dilemma experiment.

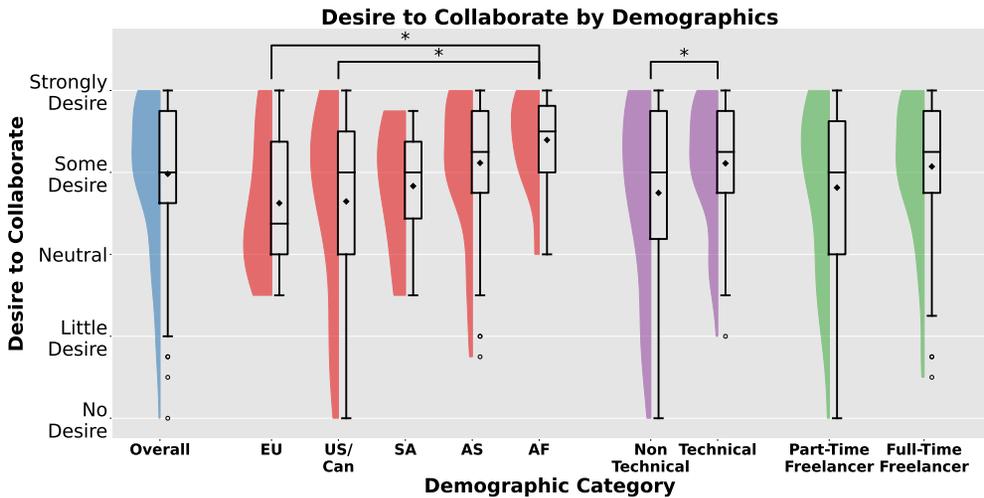

Fig. 2. Across a range of demographics we find respondents expressed a desire to participate in collaborative work. But desire does significantly vary by continent and type of work. ∗ indicate significant pairwise difference of at least $p < 0.05$, all others are not significantly different.

## 4.1 Freelancers Want to Collaborate

We find strong evidence that the majority of freelancers are interested in collaborative work. In fact, *82% of respondents held a net positive view of participating in collaborative work with other freelancers.*[1] The desire to collaborate is widespread and consistent across freelancers of different demographic backgrounds (Fig. 2). For example, we find no significant difference in desire to collaborate between part-time vs. full-time freelancers (measured as earning the majority of their income on freelance platforms; t-test, $p = 0.13$), between worker quality (measured as "Top Rated" achievement badge; t-test, $p = 0.51$), or between educational attainment (measured as bachelor's degree or less; t-test, $p = 0.94$).

We do find significant differences across a few notable categories. The mean desire to collaborate for respondents from Africa was significantly higher than that of those living in Europe (t-test, $p = 0.02$) and North America (t-test, $p = 0.01$). The stronger desire for collaboration in Africa could be explained by a preference for crowd work driven by greater economic opportunity rather than the greater autonomy of remote work. These freelancers may miss the cooperative and social aspects of traditional employment. Another key demographic distinction is that the desire for collaboration was higher for technical than non-technical freelancers (t-test, $p = 0.03$). Despite overall interest in collaboration, many freelancers also hold strongly negative views on potential collaborative work. This makes sense if we consider that some freelancers have sought out this type of work because they value working independently and alone. But importantly, we find evidence that many freelancers would value the opportunity to engage in collaborative work.

## 4.2 Antecedents of Cooperation are Present

Next, we examine the extent to which antecedents for successful cooperation suggested by prior work on traditional organizations are present in the Upwork crowd. We find evidence that many factors relating to cooperative work are present (Fig. 3). We look at organizational, relational, and personal factors in turn. First, organizational factors assess the structural, cultural, managerial, and procedural dimensions of the organization thought to affect cooperation. We find strong sentiments of intrinsic satisfaction (95% agree)[2] and extrinsic satisfaction (88% agree). Generally, these factors assess the perceived quality of the worker experience, and the level of compensation for this work respectively. Extrinsic satisfaction may be particularly important as workers who feel they are not being compensated fairly may be more likely to engage in selfish behaviors. Second, we look at relational factors of cooperation. Relational factors relate to the value workers are likely to place on developing and maintaining beneficial relationships with their coworkers. We find evidence of strong identification with the broader freelancer community (90% agree), and a strong level of trust in fellow platform workers (70% agree). This suggests that a strong identification with a shared category—being a platform freelancer—could act in similar fashion to identification with a single firm for traditional workers. We also find a majority of respondents to have a sense of shared expectations and accountability (88% agree). Finally, cooperation can also be influenced by personal factors, as some individuals are predisposed to be more cooperative than others independent of any specific context. We find the majority of respondents in our survey to have a positive measure of personal cooperativeness (80% agree). Similarly, the majority of respondents showed support for learning (94% agree) and performance goal orientations (Prove 91% agree, Avoid 67% agree). This makes sense, as successfully participating in a highly-competitive online labor market requires a desire to learn and demonstrate skills. Collectively these responses suggest that the social environment of online labor market platforms may be less hostile to cooperation than previously expected.

---

[1]This gives the percentage of respondents who on average gave responses over the midpoint of the Likert scale: strongly agree or somewhat agree.

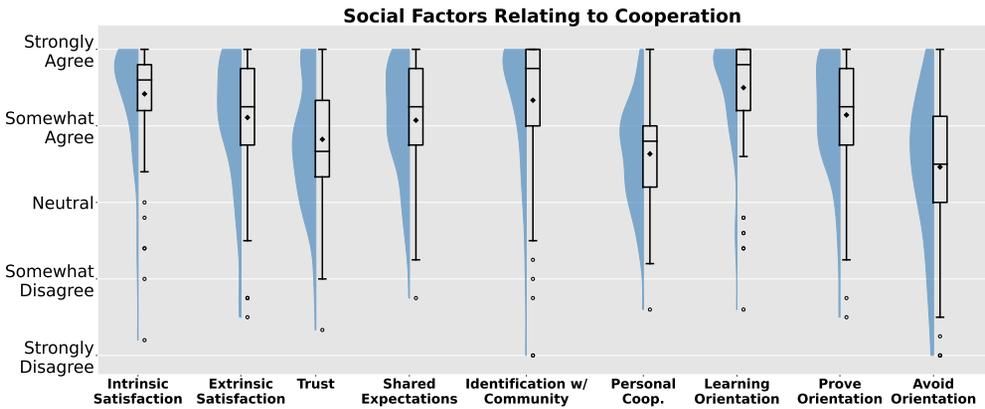

Fig. 3. Respondents expressed support for a range of factors shown to relate to cooperative behaviors.

## 4.3 Incentivized Social Dilemma

Our results provide a resounding YES to the question of whether freelancers will cooperate in a real social dilemma. In fact, we find a cooperation rate of 85%. This percentage far exceeds the average rates of cooperation seen in other comparable experiments pairing strangers in a dilemma. Previous experiments and meta-analysis have consistently found cooperation rates in a one shot prisoner's dilemma to be around 50% [12, 35, 52, 59]. Other experiments have demonstrated the effect of varied payoff structures within the one shot game. Even under the most cooperation friendly payoffs, rates of cooperation improve only modestly (60% and ~70%) [13, 14]. The highest rate of cooperation in a comparable one shot setting reported in the literature is 75% in an in-person laboratory experiment using undergraduate students [32]. It is therefore surprising that globally distributed, anonymous freelancers who have never met in person would exhibit such high rates of cooperation.

---

[2]Here, and throughout, "agree" gives the percentage of respondents who, on average across the items of a construct, responded above the midpoint of the Likert scale: "strongly agree" or "somewhat agree".

We found our rates of cooperation to be robust to the moral framing of the question [48]. In a pilot version of our survey we used "Cooperate" and "Defect" as labels to identify the choices available to respondents, while the survey for our main analysis avoided such a moral framing and used the neutral labels "A" and "B". The rate of cooperation under moral labels was 86% (N=113), nearly the same as under neutral labels. Further, to address concerns about the role of priming, in our main survey the social dilemma was the first question respondents answered, while it was the last question in our pilot survey. This suggests priming and moral framing did not play a major role in respondents' decision to cooperate.

One may wonder if the size of the incentive in the prisoner's dilemma was appropriate for the sample population. We provide three pieces of evidence that suggest that it was. First, our incentives are comparable to the magnitude of incentives used in most other dilemma experiments, especially considering we administered a single-shot game which requires only a short time commitment, rather than a repeated game [4, 13, 14]. Second, it is important to note that the $2 incentive represents substantial domestic purchasing power for a large share of our respondents. Third, as a robustness test for the effect of incentive size on cooperation in our dilemma, we regressed respondents' posted hourly rate on their decision to cooperate while including other relevant demographic controls such as age, education, continent and technical background. None of the demographic controls, nor posted hourly rate ($p = 0.35$) were significant. This suggests that respondents who earn lower wages, and therefore face relatively larger incentives in our dilemma, were not meaningfully less cooperative. This is suggestive that the actual size of the incentive—despite obvious differences in local purchasing power—does not appear to affect our results.

### 4.3.1 Freelancers Perceive Fellow Freelancers to have Cooperative Intentions.

*4.3.1 Freelancers Perceive Fellow Freelancers to have Cooperative Intentions.* An individual's perception of how others will behave is often an important driver of how they actually behave [9]. Therefore, we collected data not only on the choices made in our social dilemma but also on respondents' prediction of the choices others would make [14]. As expected, the predicted rates of cooperation were higher for those who choose to cooperate rather than defect, 68% vs. 44%. Overall, respondents predicted that 64% of fellow freelancers would choose to cooperate (when 85% did in fact cooperate). Only 16% of respondents predicted less than half of respondents to the survey would also cooperate. We find that the strength of a freelancer's identification with the freelance community is not predictive of their dilemma choice, while their predicted rate of cooperation by others is highly predictive of their own choice.

## 4.4 Mixed Preferences for How to Organize Collaborative Crowd Work

Given prior work in the CSCW community comparing workflow and role-based systems for crowdsourcing complex work [53] and the interest in building crowdsourcing systems [39, 54, 61, 67], we asked respondents how they would like collaborative work to be organized. Respondents did not express a clear preference for the use of workflows versus the use of clearly defined roles to coordinate a hypothetical collaborative task (54% preferred roles).

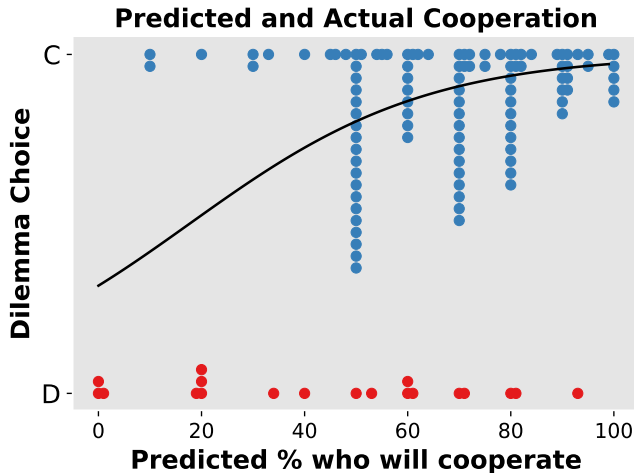

Fig. 4. The large majority of respondents believed at least half of fellow respondents would select the cooperative choice in the dilemma. Red points represent the predictions of those who choose to defect, and blue points represent the predictions of those who choose to cooperate in the dilemma. We find a positive relationship between the predicted cooperation of others, and likelihood of choosing to cooperate oneself (given by the plotted logistic regression).

## 5   MODELING RESULTS AND INSIGHTS

How could successful collaboration in online labor markets and among freelancers more generally be established? Reliable cooperation is the most important challenge any platform trying to facilitate team-based crowd work would need to solve. We explore the likely mechanism through which collaboration can be facilitated in online labor markets through a structural equation model with sequential mediation (Fig. 5) and suggest a potential pathway for platforms to strengthen cooperation. We find a significant ($p = 0.048$) indirect pathway linking a sense of shared expectations and accountability to cooperation in the dilemma through trust in fellow freelancers as a mediator, which is itself mediated by the expectation of cooperation.

We find that the expected rate of cooperation by others mediates the relationship between trust in fellow freelancers and actual cooperation ($p = 0.023$). This result is in line with previous research that has emphasized the role of expectations of others in determining behavior [9], and the role of trust in the formation of positive expectations. Therefore, this result confirms, unsurprisingly, that the facilitation and maintenance of trust in fellow freelancers is an essential way for platforms to strengthen cooperative behavior.

Trust in fellow freelancers, however, is a nebulous concept lacking clear actionable design recommendations for a platform operator. What is needed is a more actionable pathway that can influence the formation of trust in fellow freelancers. Our model provides strong evidence that one such mechanism is a sense of a shared sense of expectations and accountability among freelancers. This set of questions asks freelancers if they feel a shared sense of what is expected/acceptable behavior by freelancers and project requester, and if freelancers who violate these norms will be held accountable. We find that trust in fellow freelancers mediates the relationship between shared expectations and accountability and the predicted rate of cooperation ($p = 0.006$). Together, these results provide a sequentially mediated pathway that can inform the facilitation of cooperation among freelancers.

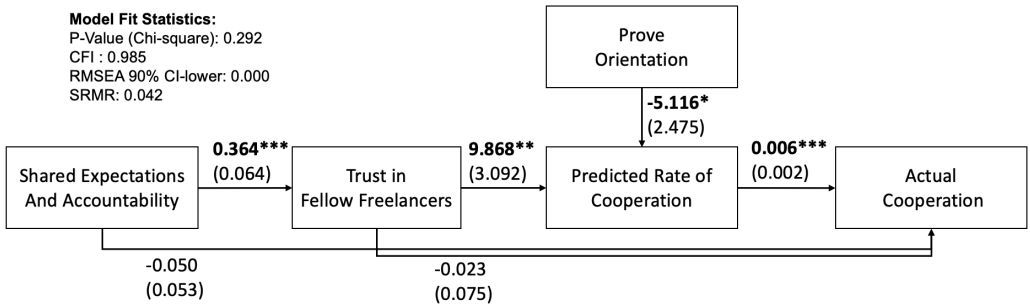

Note. Model reports unstandardized regression coefficients and significance levels based on sequential structural equation modeling using reweighted response data and 5,000 bootstrap simulations. Standardized regression coefficients in parentheses. ***p < 0.001, **p < 0.01, *p < 0.05

Fig. 5. Our SEM demonstrates a positive relationship between shared expectations and accountability, and actual cooperation through multiple mediation steps. In this case the magnitude of indirect effect is the product of the respective regression path coefficients [69]. The output also suggests strong model fit on the basis of four common measures [40], Chi-square $p > 0.05$, CFI > 0.9, RMSEA 90% CI-lower < 0.05, and SRMR < 0.05. The unweighted SEM also finds a significant sequentially mediated pathway ($p = 0.038$).

## 6   DISCUSSION

We find the freelancers in our sample show strong signs of belonging to a coherent group, and a desire—and willingness—to cooperate. Respondents also expressed feelings of shared expectations and trust in fellow freelancers. Together, our results suggest that rather than behaving as disconnected anonymous participants in a transaction-based market, freelancers instead show signs of a shared culture. One may wonder how generalizable those results are to the broader population of "gig" and crowd workers. While our sample is comprised of freelancers on Upwork specifically, we expect positive feelings towards collaboration to generalize most readily to other online labor markets focused on knowledge work (higher levels of skill use, longer term engagement, more complex tasks). Our results may generalize less readily to online labor markets focused on microtasks or capital-intensive work (e.g., Uber drivers). However, the potential for collaboration is greatest and most relevant among knowledge workers [56].

Our results suggest that if designed properly, successful and reliable collaboration on online labor market platforms supporting complex knowledge-based work may be possible. Our study primarily contributes to two broad areas of theory. First, our work complements prior ethnographic work on freelancers by approaching the problem from a different methodological (and in part theoretical) standpoint. Namely, we collect real behavioral responses and focus on a relational perspective to understand how these freelancers feel about their fellow platform freelancers, the possibility of working with one another, and how they may behave in interdependent settings [11, 23, 29, 38, 64]. Those insights help inform how platforms can be designed to strengthen and cultivate the social conditions supportive for collaboration. Second, our work relates to understanding the boundary of the firm in the context of modern crowd work. This theory considers what types of problems are better solved within firms compared to transacted on a markets, and why this is the case [72]. The respondents to our survey are not members of the same firm, but are drawn from a large, globally distributed marketplace on which they rarely interact. Yet we see evidence of thick social properties such as trust, identification, and cooperation that were previously thought to only occur within firms. This informs not only how we should think about freelancers, but also how managers and policy makers should contend with simultaneous contributions from internal and external (freelance) workers [37].

Given the unusually high rate of cooperation in our survey, the natural question to ask is why? None of Deutsch's categories of cooperative intentions (personal affection, exchange, third-party, and conscious intentions) can be seen as applying to our scenario [21]. Therefore, we suggest that a good explanation for the high rates of cooperation may be the presence of an ingroup cooperation bias.[3] This hypothesis is supported by our finding of strong identification with the broader freelancer community. Theory suggests that individuals are more willing to incur a personal cost to benefit ingroup members, compared to outgroup members. A meta-analysis has confirmed this effect is strongest in situations involving interdependent outcomes such as our social dilemma [3]. This provides further evidence freelancers on Upwork are not simply actors of a purely transaction-based labor market but display elements of social cohesion that are more consistent with traditional work arrangements within firms.

Our results also indicate that most respondents believed the majority of fellow freelancers completing the survey would also choose to cooperate. Yet, fellow freelancers represents a collection of nameless unspecified strangers. Therefore this outcome suggests the presence of depersonalized trust. Namely, the general expectation that others will be cooperative within an ingroup. Depersonalized, or swift trust, plays an essential role in facilitating initial cooperation in new teams until it can be replaced by knowledge-based forms of trust developed through previous interactions [47, 58]. This shows that the shared category of being a freelancer on the same platform might provide an initial baseline level of depersonalized trust. There is little reason to think that such platform-based identification would be different on Upwork than on other platforms, even if the nature of tasks is different, as indicated by signs of helping behavior like sharing best practices for microtask work [29]. Future research should examine the extent of this effect, and if it alone is sufficient to consistently enable initial cooperation in real work settings, or if other steps must be taken to enhance the formation of swift trust in crowd teams.

Respondent's expectation that other respondents would also choose to cooperate can also provide insights into the mechanism driving cooperative behavior. Two mechanisms that have been suggested to cause ingroup cooperation are the social identity theory, and the bounded generalized reciprocity hypothesis [49]. We find that the strength of one's identification with the freelancer community is not predictive of cooperation, while the expected rate of cooperation by others is highly predictive of the choice to cooperate. Therefore, we theorize that a general expectation of reciprocity, and not a desire to maintain or enhance the feeling of group membership, drove cooperation. That is, freelancers cooperated because they expected others would do the same, not because they placed a high value on enhancing their sense of connection with other respondents. This would define both an opportunity and a limitation of a platform like Upwork: despite being globally distributed, interacting mostly through digital channels, and so far having few opportunities to interact with each other, freelancers already have enough in common to elicit an ingroup effect. However, it appears they they do not have *so* much in common that they desire to develop this sense of group beyond practical considerations.

The mediation pathway we explore suggests that a sense of shared expectations, and accountability for those who deviate from these expectations, is significantly linked to actual cooperation. This is consistent with existing literature on the functioning of (in-person) temporary teams. Repeated social reinforcement of expectations, or preexisting occupational norms, can provide a shared sense of acceptable and desired behavior that aids the functioning of temporary teams [5, 6]. The perception that authorities can provide accountability to those who engage in harmful behaviors has also been shown to play an essential role in the emergence of cooperative behaviors in temporary teams [65]. Further, the sense of shared expectations and accountability we assessed, have been formed in the context of individual rather than cooperative work, as the platform does not currently support collaboration. We do not yet understand what norms and expectations will exist in team-based crowd work given the novelty of this form of work [37], and the changing composition of worker expertise.

---

[3]Interestingly, intrinsic and extrinsic satisfaction are *not* predictive of desire to cooperate. This suggests that respondents are not simply willing to cooperate because they feel good and satisfied, further supporting our suggestion that the high rate of cooperation may be driven by an ingroup effect.

### 6.1 Implications for Platform Design

The mediation modeling results of this study provide evidence that theories of coordination and cooperation developed for in-person temporary teams [5, 65, 66] may be usefully extended to (temporary) crowd teams. As suggested by previous ethnographic work, we find that a sense of shared understanding of expected behavior, and the expectation of accountability are highly related to cooperative behavior. In film crews, expected behaviors are frequently reinforced through dialogue [5]. Analogously, online labor platforms could reinforce expected behaviors by facilitating, encouraging, and publicizing discussions of norms and expectations within the community, and within the context of a specific project team. In emergency departments, accountability can be enhanced through a task allocation system that eliminates the potential benefits of reducing effort [65]. This gives workers confidence that authorities are acting to prevent cheating behaviors. Translating a strong sense of accountability to crowd teams is a challenging and open question. A peer-based posthoc rating system would do little to address freelancers' concerns about cheating behaviors during the period in which a project is being completed, and threats of negative retaliatory ratings could undermine its effectiveness. This may lead to cases where a single freelancer engaging in uncooperative behavior could lead to the deterioration of cooperation across the entire team. We therefore hypothesize that platforms must adopt some form of real-time accountability and dispute resolution aided by algorithmic governance [36].

### 6.2 Implications for Management of Workforce Ecosystems

As organizations become increasingly reliant on blending internal (i.e., permanent) and external (i.e., temporary or "gig") contributors [37], managers must contend with the increased complexity of overseeing this process. Managerial practices must be adapted to address key questions such as how work should be allocated to internal vs. external contributors. Managers struggling with these questions are likely to assume a lack of shared norms and culture for freelancers as a collective group. Our research suggests this is not necessarily the case, possibly changing how these decisions should be made. Existing logic would dictate that any task with significant interdependencies should be handled entirely internally. The development of effective team-based crowd work may provide an alternative option to managers.

Beyond interdependent tasks, our work has implications for how managers should think about the cultural contexts on either side of their firm boundary. Because managers are aware of the norms and expectations within their organization, they have less uncertainty about potential outcomes. However, if this understanding does not extend to external workers, it may create a desire to assimilate them into aspects of internal culture. In the future, managers may alternatively seek to understand the collective culture and norms of their external workforce, thereby gaining insight into the expectations of these workers [2]. One can imagine a world in which different labor platforms have centered around unique collective norms and expectations much like firms today. Rather than contributing to a porous boundary of the firm [72], managers who recruit freelancers from online labor markets may eventually be better conceptualized as quasi-firms in their own right. This contributes the evolving sense that much work on theories of the firm in the digital age [46] does not go far enough in recognizing and considering how digital technologies and the increasing use of AI could fundamentally reshape the way work is organized. Often they offer only a short-term view of the firm in the digital age that is too reliant on the existing status quo. Ultimately, as with most technologies, the digital age firm will be defined by the adoption of entirely new processes only enabled by digital technologies, rather than the digital augmentation of existing processes.

### 6.3 Limitations

Our study is not without limitations. First, Upwork is a large platform with millions of freelancers, so our survey may not be representative of the perspectives of all Upwork freelancers. Second, our sample of respondents come from a single platform (Upwork). Consequently, we cannot make cross-platform comparisons and our conclusions may not generalize to other platforms. Compared to other online labor markets, Upwork is attracting freelancers for knowledge-based tasks that are slightly more complex and slightly more long-term than on other platforms. For example, Amazon Mechanical Turk (AMT) is specialized on very small tasks ("microtasks"). Future work should investigate other platforms such as AMT and those facilitating in-person services such as carework platforms. Third, our dilemma used incentivized outcomes ranging between $0 to $2 of bonus payment. While we find no indication that the payment was too small, and it is consistent with payments used in prior work, future work could investigate higher incentives. Fourth, consistent with past organizational research, we use cooperation in an incentivized dilemma as a proxy for real cooperative behaviors [18]. Future work could more directly measure cooperative behaviors during real-effort collaborative projects.

## 7 CONCLUSION

This paper indicates that many freelancers desire to collaborate with other freelancers, are willing to cooperate with each other, and display traits consistent with belonging to a coherent group or organization. Specifically, we suggest an ingroup cooperation bias may best explain our key results. Finally, in accordance with research on in-person temporary teams, we find that a sense of shared expectations and accountability is an important predictor of cooperation. These findings have implications for platforms seeking to facilitate team-based crowd work, and for organizations that rely on external contributors. We point to initial suggestions and important future research for the design of platforms facilitating online collaboration between strangers. We also caution managers to not incorrectly presume a lack of shared culture among freelancers. We support the emerging view that interactions on labor platforms can and should be relationally focused, collaborative, and helpful [51]. Overall, it appears that some platform crowds are much more social than previously thought [24, 29, 55].

## ACKNOWLEDGMENTS

Melissa Valentine and Hatim Rahman for helpful comments on the manuscript.

## A  SURVEY QUESTIONS

Questions 1 and 13 require respondents to choose a single answer, question 2 users a sliding button representing integer choices between 0 and 100, all other questions use a 5-point likert scale with 1 = Strongly Disagree, 5 = Strongly Agree.

(1) **Social Dilemma:** (labels not included in survey)

In this task you will be paired with another freelancer on Upwork. Your response will remain confidential. You may choose option A or option B. This choice will yield one of the payments below which you will receive as a bonus upon completion of the survey.

If you choose A and they choose A you get $1.50 and they get $1.50.

If you choose A and they choose B you get $0.00 and they get $2.00.

If you choose B and they choose A you get $2.00 and they get $0.00.

If you choose B and they choose B you get $0.50 and they get $0.50.

- A
- B

(2) **Predicted Cooperation**

What percentage of respondents do you believe will choose option A?

- Discrete slider from 0 to 100

(3) **Intrinsic Satisfaction**

I am satisfied with...

- How interesting my work is on Upwork.
- Working on Upwork makes good use of my abilities.

- The degree of freedom I have when working on Upwork.
- The feelings of accomplishment I get from working on Upwork.
- The opportunities working on Upwork provides me to interact with others.

(4) **Extrinsic Satisfaction**
I am satisfied with…
- The compensation I receive working on Upwork.
- The fairness of my earnings in relation to efforts I expend.
- My probable future earnings on Upwork.
- The public perception and attitude towards freelance work.

(5) **Trust in Fellow Freelancers**
I consider my fellow freelancers on the platform to be people who…
- Have high integrity.
- Can be trusted completely.
- Can be counted on to get the job done right.
- Cannot be trusted at times.

(6) **Shared Norms and Accountability**
I believe…
- Freelancers who engage in unethical/misleading behaviors will be held accountable.
- Freelancers have a shared sense of what is acceptable/expected behavior in regards to the completion of their tasks.
- Freelancers have a shared sense of what is acceptable/expected behavior from requestors.

(7) **Identification with Community**
These questions are about your feelings towards the Upwork freelancer community.
- I am proud to think of myself as a member of the freelancer community.
- I would tell my friends freelancing on Upwork is a great work opportunity.
- I feel loyal to the freelancer community.
- I am proud to tell others I freelance on Upwork.

(8) **Personal Cooperativeness**
These questions are about how you see yourself.
- I can be a good team player.
- I enjoy activities that involve a high level of cooperation with other people.
- I prefer to work independently rather than in a group.
- I find more satisfaction working toward a common group goal than working toward my individual goals.
- I find joint projects with other people very satisfying.

(9) **Learning Goal Orientation**
These questions are about your preferences.
- I am willing to select a challenging work assignment that I can learn a lot from.
- I often look for opportunities to develop new skills and knowledge.
- I enjoy challenging and difficult tasks where I'll learn new skills.
- For me, development of my work ability is important enough to take risks.

• I prefer to work in situations that require a high level of ability and talent.

(10) **Prove (Performance) Orientation**

These questions are about your preferences.
• I'm concerned with showing that I can perform better than other freelancers.
• I try to gure out what it takes to prove my ability and skills.
• I enjoy it when others are aware of how well I am doing.
• I prefer to work on tasks where I can prove my ability to others.

(11) **Avoid (Performance) Orientation**

These questions are about your preferences.
• I would avoid taking on a new task if there was a chance I would appear incompetent.
• Avoiding showing a weakness is more important to me than learning a new skill.
• I prefer to avoid tasks where I might perform poorly.
• I'm concerned about taking on a task if my performance would reveal that I have low ability.

(12) **Desire to Collaborate**

Assuming fair payment and effective communication tools, how do you feel about the following statements?
• I would like to work jointly with one or more freelancers on some jobs.
• I would like to collaborate with other freelancers in order to complete larger or more difficult tasks.
• By sometimes working with other freelancers I would be able to complete more diverse and interesting tasks.
• The option to work jointly with other freelancers would increase my overall work satisfaction.

(13) **Workflow vs. Roles**

If you were completing a large task jointly with one or more freelancers which scenario would you prefer?
• The requestor provides a plan of what subtasks need to be completed, in what order they should be completed, and who should complete each subtask.
• The requestor communicates what role each freelancer has, and you and your fellow freelancers can decide how to best organize the work.